\documentclass{csmagclass}														
\usepackage{graphicx}
\usepackage{subfigure}	
\begin{document}		
\headings																															 %

\title{{ Spin-glass-like ordering in a frustrated $J_1-J_2$ Ising antiferromagnet on a honeycomb lattice }}
\author[1]{{ M. \v{Z}UKOVI\v{C}\thanks{Corresponding author: milan.zukovic@upjs.sk}}}
\author[1]{{ M. BOROVSK\'{Y}}}
\author[1]{{ A. BOB\'{A}K}}
\author[2]{{ T. BALCERZAK}}
\author[2]{{ K. SZA{\L}OWSKI}}
\affil[1]{{Institute of Physics, Faculty of Science, P. J. \v{S}af\'arik University in Ko\v{s}ice, Park Angelinum 9, 041 54 Ko\v{s}ice, Slovakia}}
\affil[2]{{Department of Solid State Physics, University of \L\'{o}d\'{z}, Pomorska 149/153, 90-236 \L\'{o}d\'{z}, Poland}}

\maketitle

\begin{Abs}
We study the nature of a low-temperature phase in the frustrated honeycomb-lattice Ising model with first- and second-neighbor antiferromagnetic (AF) interactions, $J_1$ and $J_2$, respectively, for $R = J_2/J_1 > 1/4$. It is known that for $R < 1/4$ there is a phase transition at low temperatures to the AF phase. Nevertheless, little is known about the critical behavior of the model for $R > 1/4$, except for recent effective field results which detected no phase transition down to zero temperature. Our Monte Carlo results suggest that for $R > 1/4$ there is at least one peculiar phase transition accompanied by a spin-glass-like freezing to a highly degenerate state consisting of frozen domains with stripe-type AF ordering separated by zero-energy domain walls. In spite of the local ordering within the respective domains there is no ordering among them and thus, unlike in the corresponding square-lattice model with $R > 1/2$, there is no conventional magnetic long-range ordering spanning the entire system. 
\end{Abs}
\section{Introduction}
A frustrated $J_1-J_2$ Ising model with nearest-neighbor (NN) and next-nearest-neighbor (NNN) antiferromagnetic (AF) interactions on a square lattice has attracted a lot of attention due to a long-standing controversy regarding the nature of its critical behavior. While the transition to the AF (N\'{e}el) phase for $R = J_2/J_1 < 1/2$ is believed to belong to the Ising universality class, conflicting results have been reported regarding the transition to the striped or superantiferromagnetic (SAF) phase for $R > 1/2$. A scenario proposed by a series of earlier studies suggested a second order transition with non-universal exponents for any $R > 1/2$ (see Ref.~\cite{mala06} and references within), while some more recent approaches favored a first order transition for $1/2 < R < R^*$ and a continuous one only for $R > R^*$~\cite{mora93,anjo08,kalz08,kalz11,jin12}. Nevertheless, even the value of $R^*$ in the latter cases has been a subject of controversy with rather different estimates ranging from the early $1.1$~\cite{mora93} down to the latest $0.67$~\cite{jin12}.

Much less attention has been paid to this model on a honeycomb lattice~\cite{kudo76,kats86,boba16}. A recent study within an effective field theory predicted the existence of the AF phase for $R < 1/4$, with a tricritical behavior, but no long-range ordering was found for $R > 1/4$~\cite{boba16}. Therefore, in the present study we focus on the region of $R > 1/4$ and explore the possibility of some kind of ordering at low-temperatures using a Monte Carlo approach. We find that the critical behavior in this region of the exchange parameter space is much different from both the EFT prediction of no phase transition as well as from the SAF long-range ordering observed in the corresponding model on a square lattice. In particular, at sufficiently low temperatures we observe a phase transition to a highly degenerate state consisting of frozen domains with the stripe-type AF ordering inside the domains separated by zero-energy domain walls. Consequently, the ground state appears to show no conventional long-range ordering like the one observed in the square-lattice system.
\section{Model and method}
The honeycomb-lattice Ising antiferromagnet with NN $(J_1 < 0)$ and NNN $(J_2 < 0)$ interactions can be described by the Hamiltonian 
\begin{equation}
\label{hamiltonan}
H = -J_1\sum_{\langle i,j\rangle}s_i s_j - J_2 \sum_{\langle i,k\rangle}s_i s_{k},
\end{equation}
where $s_i = \pm 1$ is the Ising spin variable at the $i$th site and the summations $\langle i,j\rangle$ and $\langle i,k\rangle$ run over all NN and NNN spin pairs.

The model is studied by standard Monte Carlo (MC) simulations, using the Metropolis algorithm and system sizes $N=L \times L$, where $L=12,24,36$, and $72$, with the periodic boundary conditions. For thermal averaging of the calculated thermodynamic quantities at each value of the reduced temperature $k_B T/|J_1|$ we consider up to $10^6$ Monte Carlo sweeps (MCS) after discarding initial twenty percent of that number for securing thermal equilibrium. The simulations start from high temperatures, using random initial configurations, and then the temperature is gradually lowered and the simulations start from the final configuration obtained at the previous temperature. Such an approach ensures that the system is maintained close to the equilibrium in the entire temperature range and considerably shortens thermalization periods.

To detect anomalies related to possible phase transitions we calculate the internal energy per spin $E/N|J_1|=\langle H \rangle/N|J_1|$ and the specific heat obtained from the energy fluctuations as
\begin{equation}
\label{spec_heat}
C/N=\frac{\langle H^{2} \rangle - \langle H \rangle^{2}}{Nk_{B}T^{2}},
\end{equation}
where $\langle \cdots \rangle$ denotes the thermal average.
\section{Results and discussion}
Following Ref.~\cite{boba16}, the ground-state energy in the expected superantiferromagnetic (SAF) phase, for $R > 1/4$, is given by $E_{SAF}/N|J_1| = - 1/2 - R$. Such energy corresponds to any spin arrangement in which the central spin has out of the three NN spins two with antiparallel and one with parallel orientation, i.e. $(2\times\uparrow\downarrow) + (1\times\uparrow\uparrow)$ and out of the six NNN spins four with antiparallel and two with parallel orientation, i.e. $(4\times\uparrow\downarrow) + (2\times\uparrow\uparrow)$. Such arrangements include the the collinear striped states, as illustrated in Fig.1 of Ref.~\cite{boba16}. However, unlike in the square-lattice case, in which there are four different ways of the arrangement, in the honeycomb lattice such states are macroscopically degenerate. 

\begin{figure}[h!]
\subfigure{\includegraphics[width=0.5\columnwidth]{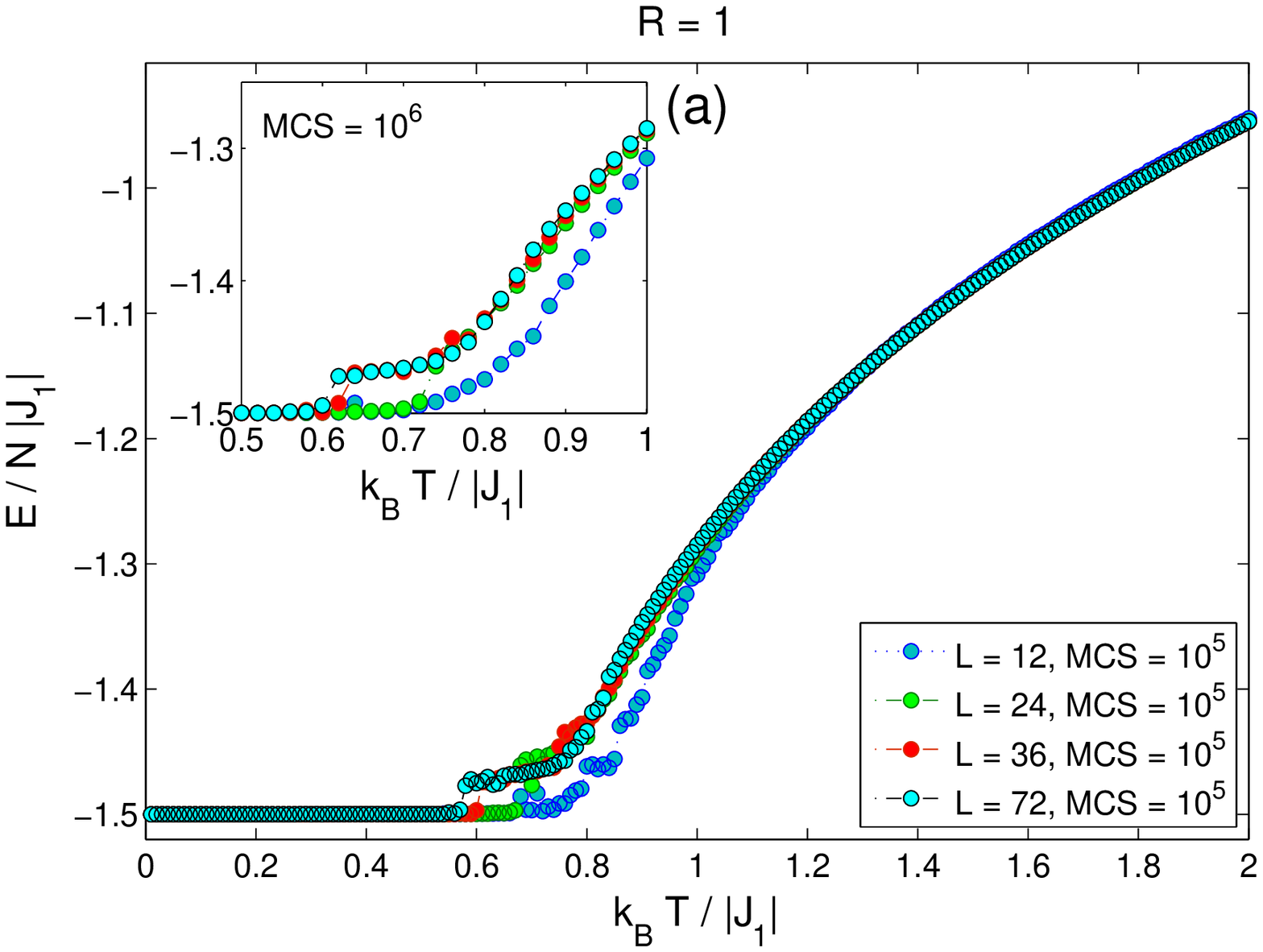}\label{fig:E_R-1}}
\subfigure{\includegraphics[width=0.5\columnwidth]{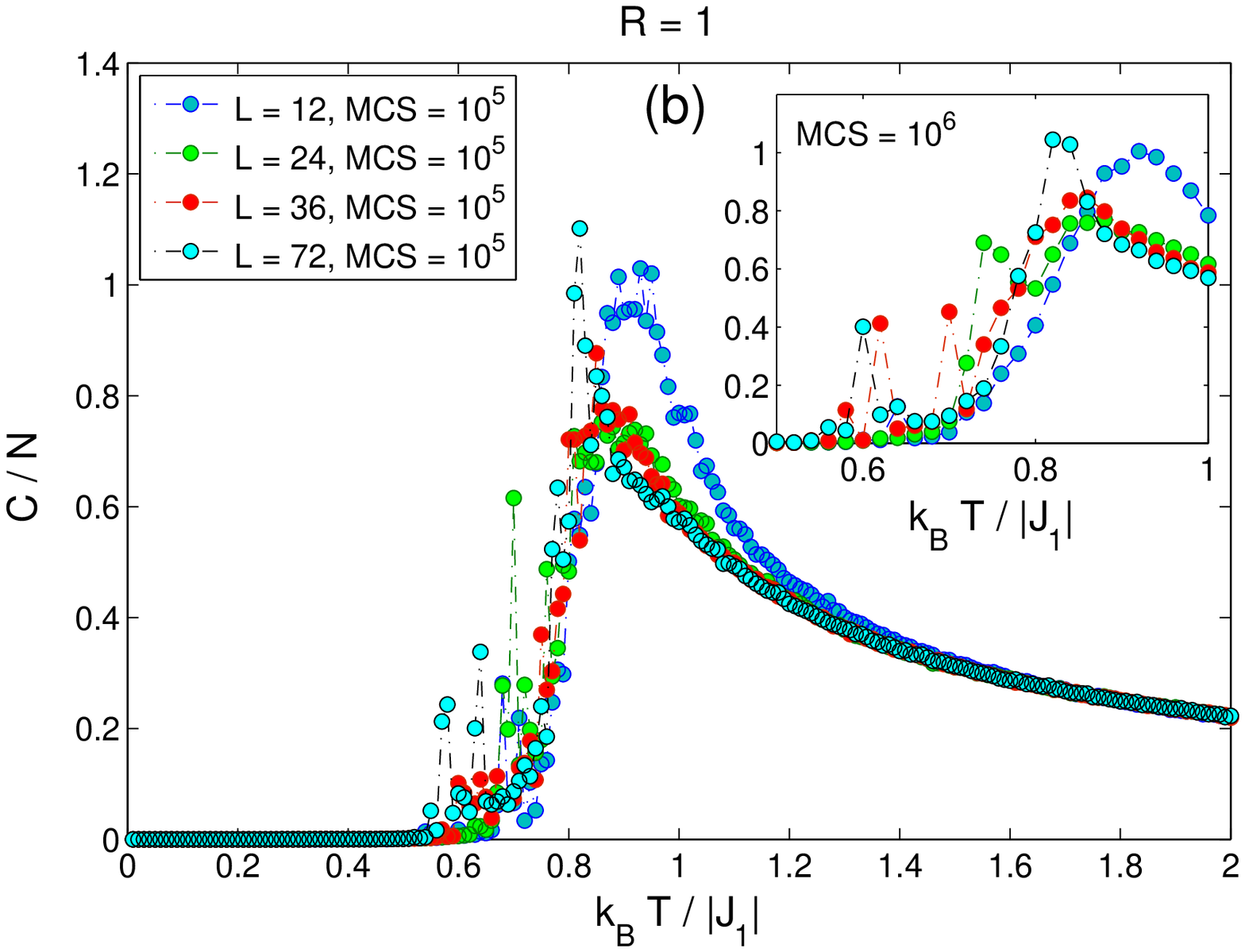}\label{fig:C_R-1}}
\caption{Temperature dependencies of (a) the internal energy and (b) the specific heat, for $R=1$ and different lattice sizes $L$. The insets show the results for $10^6$ MCS.}
\label{fig:R-1}
\end{figure}

\begin{figure}[h!]
\subfigure{\includegraphics[width=0.5\columnwidth]{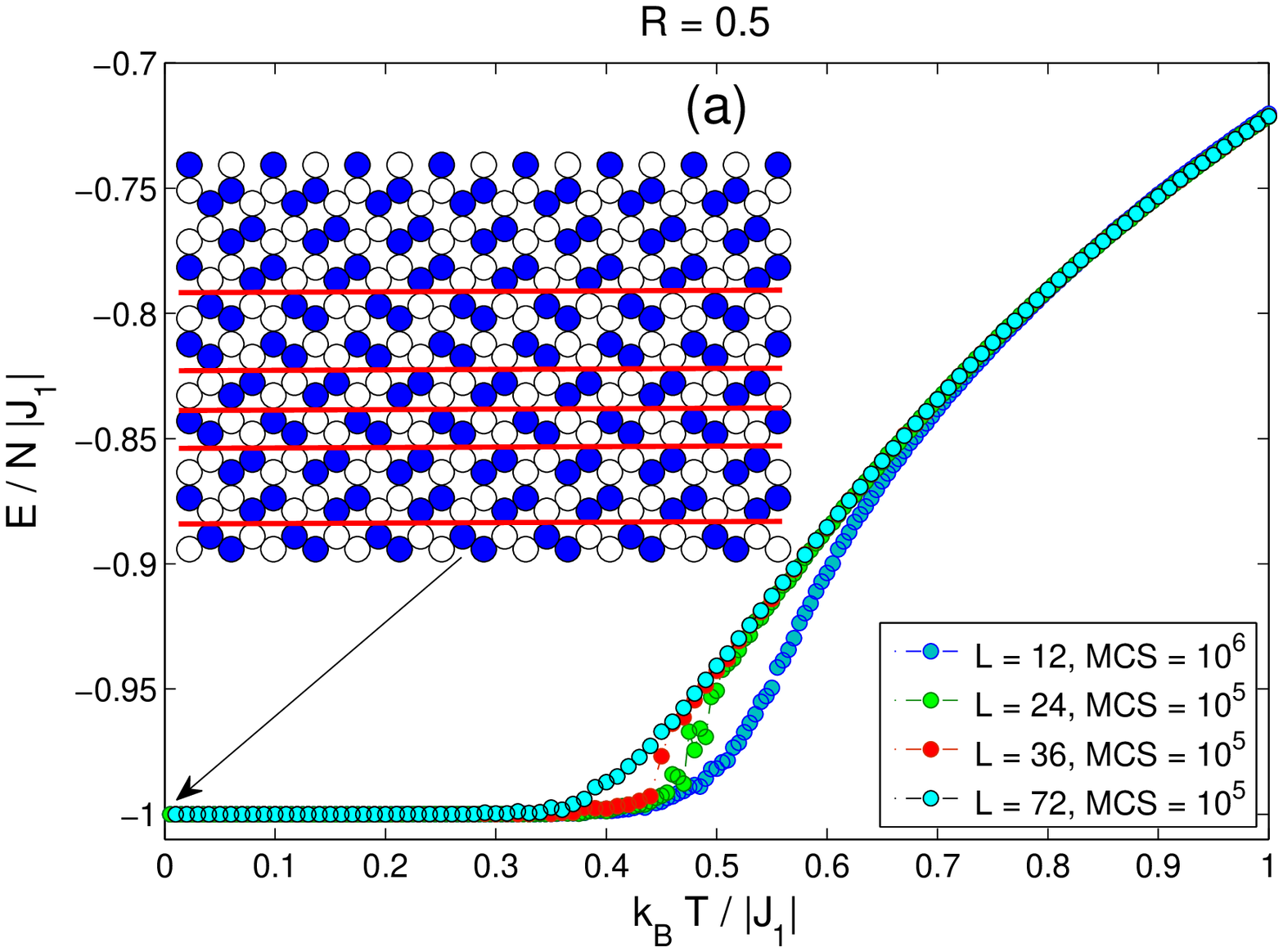}\label{fig:E_R-05}}
\subfigure{\includegraphics[width=0.5\columnwidth]{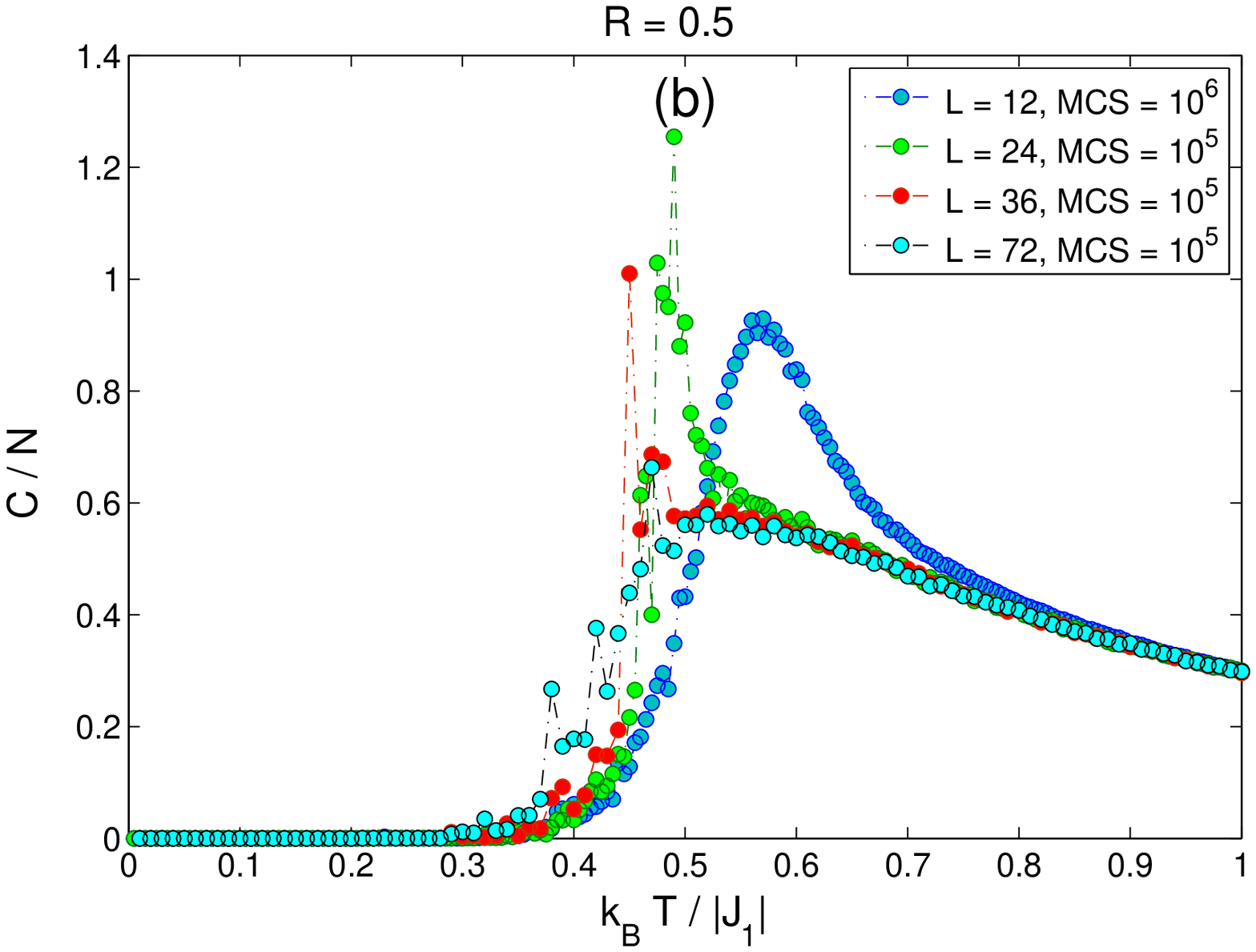}\label{fig:C_R-05}}
\caption{Temperature dependencies of (a) the internal energy and (b) the specific heat, for $R=0.5$ and different lattice sizes $L$. The inset shows a snapshot at $k_BT/|J_1| = 0.01$ for $L=36$, where the empty (filled) circles represent spins up (down) and the horizontal lines mark domain walls.}
\label{fig:R-05}
\end{figure}

\begin{figure}[h!]
\subfigure{\includegraphics[width=0.5\columnwidth]{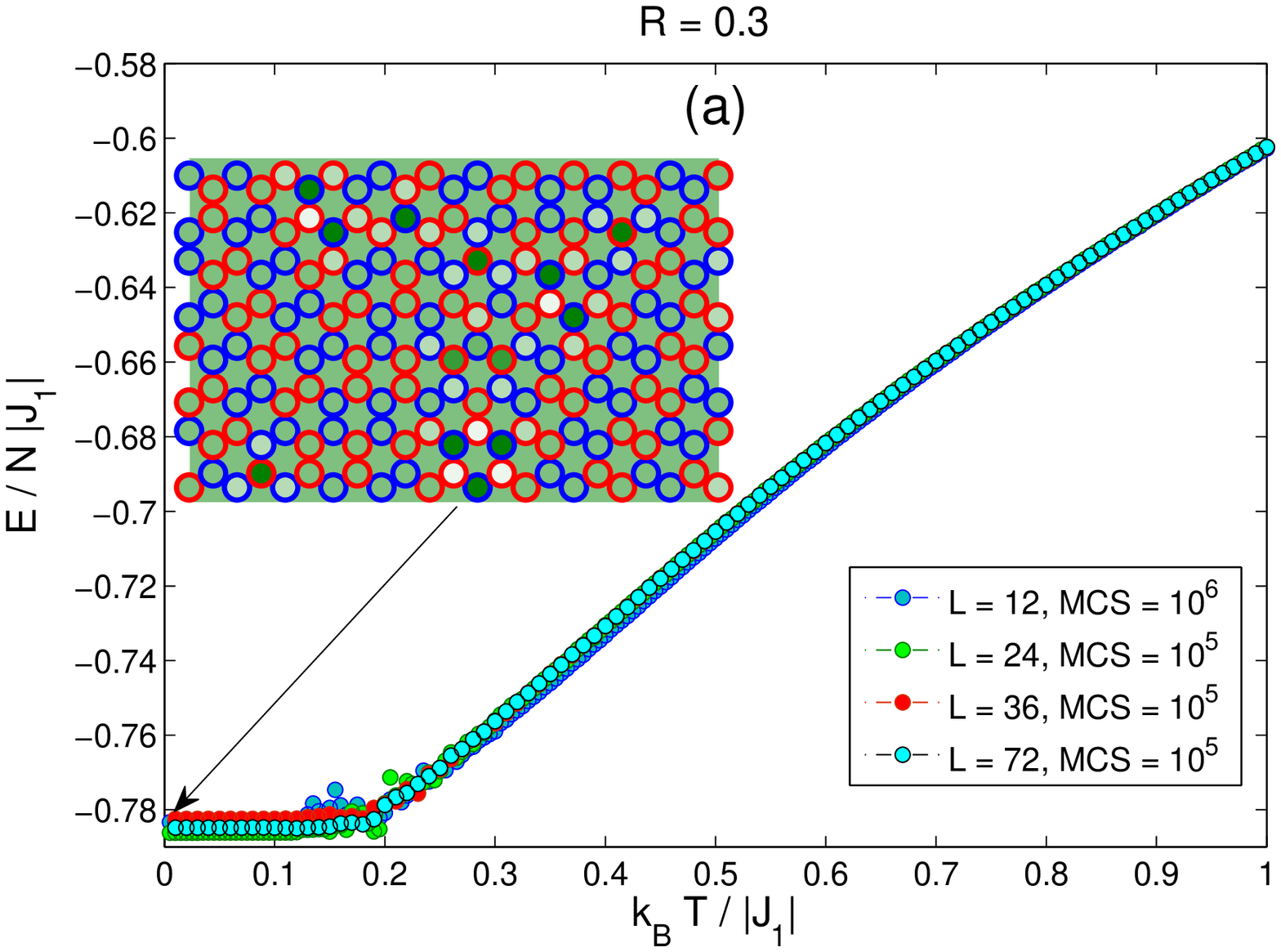}\label{fig:E_R-03}}
\subfigure{\includegraphics[width=0.5\columnwidth]{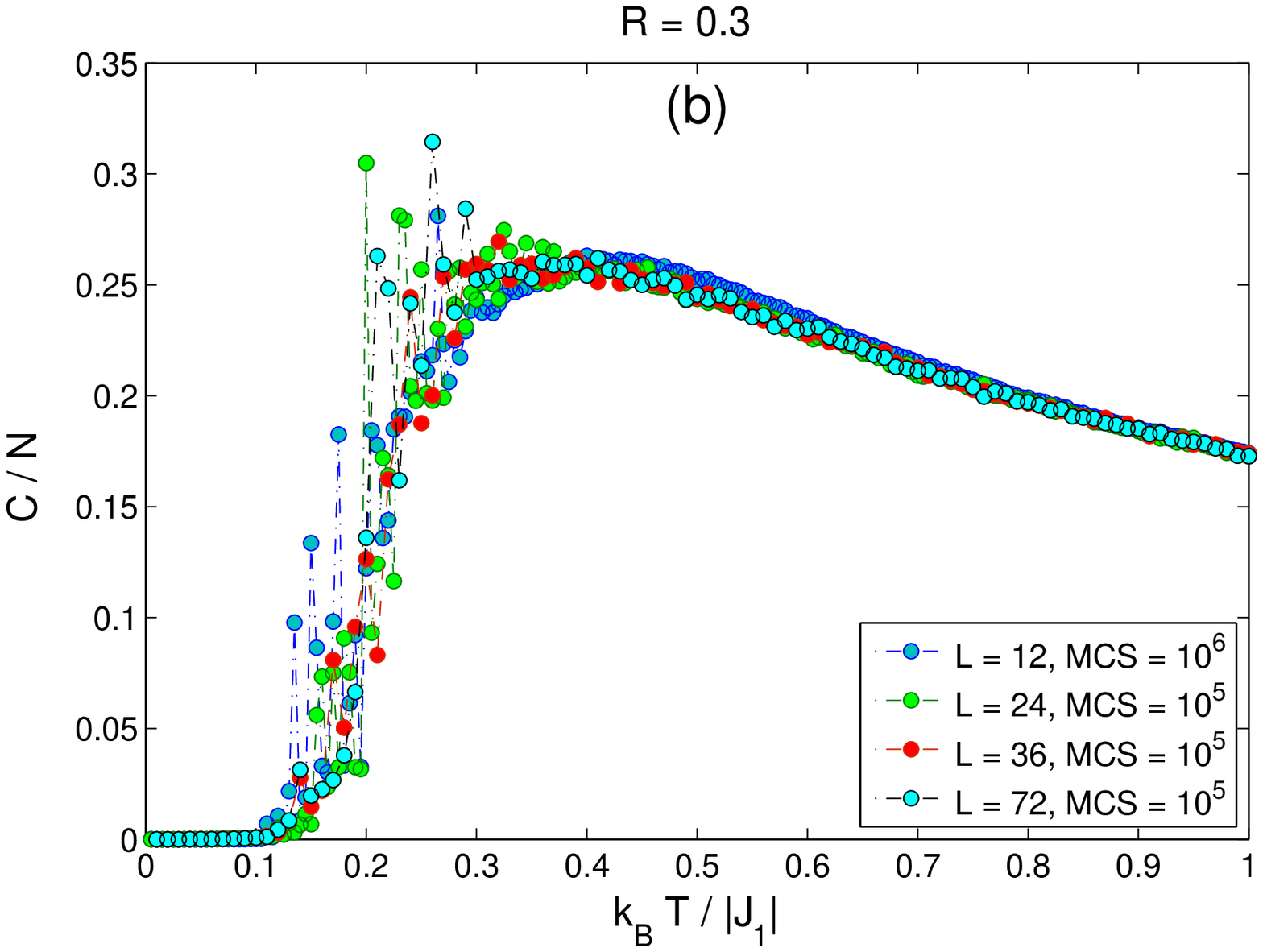}\label{fig:C_R-03}}
\caption{Temperature dependencies of (a) the internal energy and (b) the specific heat, for $R=0.3$ and different lattice sizes $L$. The inset shows a snapshot  of spins and their local energies (see text) at $k_BT/|J_1| = 0.01$ for $L=36$.}
\label{fig:R-03}
\end{figure}

Nevertheless, as will be shown below, at low temperatures there is some kind of phase transition to this highly degenerate phase with no conventional magnetic long-range ordering, which could not be detected within the effective field theory approach~\cite{boba16}. Fig.~\ref{fig:R-1} shows temperature dependencies of the internal energy and the specific heat for $R=1$. For $L=12$ the internal energy curve shows at $k_BT/|J_1|\approx 0.9$ an anomaly characteristic for a continuous phase transition reflected in a specific heat peak (Fig.~\ref{fig:C_R-1}). However, for larger $L$ there is another anomaly at lower temperatures in the form of a jump in the internal energy and the spike-like specific heat peak at $k_BT/|J_1|\approx 0.6$, resembling a first-order phase transition. The insets show that these features also persist in much longer runs using $10^6$ MCS. Just below this low-temperature transition thermal fluctuations are strongly suppressed and the system freezes in a state corresponding to the lowest-possible (ground-state) energy of $E_{SAF}/N|J_1| = -3/2$.

For $R=1/2$ the discontinuous features of the transition are much less conspicuous but some signs of their presence are still observable. The system freezes at temperatures lower than for $R=1$ and a typical snapshot at $k_BT/|J_1| = 0.01$ for $L=36$ (only a part of it is shown) is presented in the inset. One can observe that the system consists of horizontal domains of spins with SAF-type (striped) arrangement with different widths, which are separated by parallel zero-energy domain walls crossing the entire lattice (red lines). It is easy to verify that all spins have two (four) NNs (NNNs) antiparallel and one (two) NNs (NNNs) parallel and thus the local energies of the spins inside the domains and at their walls are the same.

Finally, for $R=0.3$ MC simulations for all the considered lattice sizes fail to reach the ground state. As evidenced in Fig.~\ref{fig:E_R-03}, in all the instances they freeze to the states with energies higher than the expected $E_{SAF}/N|J_1| = -0.8$. The inset shows a typical snapshot of spins and their local energies at $k_BT/|J_1| = 0.01$ for $L=36$. Spin states are shown by red (spin-up) and blue (spin-down) circles and the color of their interior represents the value of their local energy. The predominant light-green color corresponds to $E_{SAF}/N|J_1| = -0.8$, while spins filled with lighter (darker) color represent higher (lower) local energies. We assume that this spin-glass-like freezing occurs due to the fact that the transition temperatures for $R$ approaching the critical value of $R=1/4$ are lower than the freezing temperatures.
\section{Conclusions}
Our Monte Carlo simulations demonstrated that the frustrated honeycomb-lattice Ising model with second-neighbor antiferromagnetic interactions exceeding one fourth of the strength of the nearest-neighbor interaction displays at least one phase transition. The highly degenerate low-temperature phase consists of frozen SAF-like domains separated by zero-energy domain walls and lacks a conventional magnetic long-range ordering. Due to difficulties related to extremely long equilibration and autocorrelation times application of more sophisticated methods able to overcome large energy barriers in the phase space, such as the parallel tempering (replica exchange Monte Carlo), is desirable in order to better elucidate the critical behavior in this regime by studying larger systems of varying sizes and performing a finite-size analysis.
\section{Acknowledgement}
This work was supported by the Scientific Grant Agency of Ministry of Education of Slovak Republic (Grant No. 1/0531/19) and the scientific grant of the Slovak Research and Development Agency (Grant No. APVV-14-0073).
%


\end{document}